**Non-Equilibrium Heat Transport in Pt and Ru Probed by an Ultrathin Co Thermometer**

Hyejin Jang,[1*] Johannes Kimling,[1] and David G. Cahill[1*]

[1] Department of Materials Science and Engineering and Materials Research Laboratory,

University of Illinois, Urbana, IL 61801, USA

*email: hjang@berkeley.edu, d-cahill@illinois.edu

Abstract

Non-equilibrium of electrons, phonons, and magnons in metals is a fundamental phenomenon in condensed matter physics and serves as an important driver in the field of ultrafast magnetism. In this work, we demonstrate that the magnetization of a sub-nm-thick Co layer with perpendicular magnetic anisotropy can effectively serve as a thermometer to monitor non-equilibrium dynamics in adjacent metals, Pt and Ru, via time-resolved magneto-optic Kerr effect. The temperature evolutions of the Co thermometer embedded in Pt layers of different thicknesses, 6–46 nm, are adequately described by a phenomenological three temperature model with a consistent set of materials parameters. We do not observe any systematic deviations between the model and the data that can be caused by a non-thermal distribution of electronic excitations. We attribute the consistently good agreement between the model and the data to strong electron-electron interaction in Pt. By using Pt/Co/Pt and Pt/Co/Pt/Ru structures, we determine the electron-phonon coupling parameters of Pt and Ru, $g_{ep}(\text{Pt})=(6\pm1)\times10^{17}$ W m$^{-3}$ K$^{-1}$ and $g_{ep}(\text{Ru})=(9\pm2)\times10^{17}$ W m$^{-3}$ K$^{-1}$. We also find that the length scales of non-equilibrium between electrons and phonons are $l_{ep}=(\Lambda_e/g_{ep})^{1/2}\approx9$ nm for Pt and $\approx7$ nm for Ru, shorter than their optical absorption depths, 11 and 13 nm, respectively. Therefore, the optically thick Pt and Ru layers show two steps of temperature rise: The initial jump of electron temperature that occurs within 1 ps is caused by direct optical excitation and electronic heat transport within a distance $l_{ep}$ for the Co layer. The second temperature rise is caused by heat transport by electrons and phonons that are near thermal equilibrium. We contrast two-temperature modeling of heat transport in Pt an Ru films to calculations for Cu, which has a much longer non-equilibrium length scale, $l_{ep} \approx 63$ nm.

## I.  INTRODUCTION

Non-equilibrium dynamics in metals induced by ultrafast laser irradiation involve fundamental properties of heat carriers and their interactions and thus have been of great interest in condensed matter physics. In metals, electrons exclusively absorb the energy of a laser pulse and are driven out of equilibrium with other carriers, e.g., phonons and magnons. Thermalization of the metal occurs via energy transfer between the heat carriers and via thermal diffusion from the irradiated surface if the sample size is longer than the optical absorption depth. Time-resolved measurements of laser-induced non-equilibrium in metals have been widely used to investigate the electron-phonon coupling parameters of metals. [1,2]

An understanding of non-equilibrium dynamics in metals is also important in the field of ultrafast magnetism. An ultrafast laser pulse can reduce or even reverse the magnetizations of ferro- and ferrimagnetic materials on ps-timescales, [3,4] and also induce THz radiation by metallic multilayers. [5–7]. However, it is still under debate how rapidly the laser energy is distributed among different carriers and how the carriers propagate through the irradiated samples.

The laser-induced non-equilibrium dynamics are often described by a phenomenological three-temperature model (3TM) (or reduced to a 2TM for non-magnetic metals). The 3TM consists of three coupled differential equations for temperatures of electrons ($T_e$), phonons ($T_{ph}$), and magnons ($T_m$), as described by Eqs. (1–3). The energy transfer between the heat carriers $i$ and $j$ ($i, j=e, ph, m$) is described by a coupling constant, $g_{ij}$. The phonon-magnon coupling ($g_{pm}$) is often ignored in ferromagnetic metals because phonon-magnon coupling is typically much weaker than electron-magnon coupling. [8] $C_i$ and $\Lambda_i$ are the heat capacity and thermal conductivity of a carrier type $i$, respectively. The absorption of laser energy by electrons is described via a source term, $S(z,t)$, see Eq. (4). $P(t)$ is a temporal profile of a pulse intensity and $A(z)$ is the absorption profile calculated using a transfer matrix method with refractive indices of constituent materials; $S_0$ is a pre-factor to normalize $S(z,t)$ to an absorbed laser fluence. The temperature evolutions of the three types of heat carriers as a function of position and time, $T_i(z,t)$, are obtained by numerically solving Eqs. (1–3).

$$C_e \frac{\partial T_e}{\partial t} = \Lambda_e \frac{\partial^2 T_e}{\partial z^2} - g_{ep}(T_e - T_{ph}) - g_{em}(T_e - T_m) + S(z,t) \tag{1}$$

$$C_{ph} \frac{\partial T_{ph}}{\partial t} = \Lambda_{ph} \frac{\partial^2 T_{ph}}{\partial z^2} + g_{ep}(T_e - T_{ph}) \tag{2}$$

$$C_m \frac{\partial T_m}{\partial t} = \Lambda_m \frac{\partial^2 T_m}{\partial z^2} + g_{em}(T_e - T_m) \tag{3}$$

$$S(z,t) = S_0 P(t) A(z) \tag{4}$$

However, the 3TM has been criticized for its underlying assumption of internal thermal equilibrium of each type of heat carriers, i.e., a temperature can be defined for each carrier type. This assumption is not always justified, especially when electronic excitations deviate from the Fermi-Dirac distribution immediately after laser absorption. The photo-excited electrons are predominantly scattered via electron-electron interactions, i.e., with ground-state electrons at a low excitation density, $<10^{-3}$ of the total electron density, and further with other electronic excitations at a higher excitation density. [9] By solving the Boltzmann transport equations, Ref. [10] showed that the 2TM is valid if the timescale of electron-electron scattering ($\tau_{ee}$) is much shorter than the timescale of electron-phonon scattering ($\tau_{ep}$).

The non-equilibrium dynamics have been extensively studied for simple metals (e.g., Al) [11,12] and noble metals (e.g., Au, Ag, and Cu) [13–17]. For simple and noble metals, the electron band structures near the Fermi energy are dominated by *s* and *p* orbitals. For these metals, $\tau_{ee}$ is relatively long and often comparable to $\tau_{ep}$. Thus, non-thermal electrons can interact with phonons at the energy exchange rate that is different from thermal electrons. Experiments [16] and first-principles calculations [17] show that $g_{ep}$ of non-thermal electrons is smaller than $g_{ep}$ of thermal electrons for Ag, Au, and Al. Moreover, the mean-free-paths of electrons near the Fermi surface are much longer than the optical absorption depth. For example, the bulk mean-free-paths of electrons derived from the size-dependence of electrical resistivity of thin films or nanowires are ≈40 nm for Au and Cu. [18,19] Thus, hot electrons in optically thick metallic layers can move ballistically. Experimentally, the ballistic transport of hot electrons is supported by the apparent linear relationship between the thickness of metal layers and travel time [13] and the longer effective absorption depth for thermal modeling. [2]

The lifetimes of electronic excitations in transition metals are much shorter than the lifetimes in noble metals as the localized $d$ bands near the Fermi level enhances electron-electron scattering probabilities. [9] Also, the mean-free-paths of electrons are much shorter, e.g., 7-10 nm for Pt. [20] Therefore, the non-equilibrium dynamics of transition metals are more likely to agree with the 2TM. According to Ref. [10], Pt and Pd are close to meeting the criteria for the 2TM to be valid due to their strong electron-electron interactions. See Appendix A for further discussion of the lifetimes of electronic excitations in Pt.

In this work, we use the magnetization of an ultrathin layer of Co as a thermometer to study non-equilibrium dynamics in two transition metals, Pt and Ru. While the non-thermal behaviors of electrons in noble metals have been studied comprehensively, it is not clarified whether the laser-induced dynamics in transition metals, such as Pt and Ru, would be adequately described by the 3TM. Moreover, it is not straightforward to correlate the measured optical reflectance or transmittance with the temperatures of electrons and phonons while they are at different temperatures. When a temperature excursion is small, i.e. $\Delta T < 50$ K, and the sample stays far below the Curie temperature, the magnetization of a Co layer can be assumed to vary linearly with the magnon temperature. [21,22] By measuring the magnetization of a sub-nm-thick Co layer of Co via the magneto-optic Kerr effect, we can selectively monitor the magnon temperature in Co.

First, we prepare three Pt/Co/Pt trilayer samples, in which Co is embedded in the Pt layer of different thicknesses, 6–46 nm. We perform pump-probe experiments with the pump beam incident on either of the two surfaces and the probe beam incident on the surface closer to Co. By varying the relative position of the Co thermometer with respect to the irradiated Pt surface, we examine transport of carriers as well as electron-phonon coupling of Pt, $g_{ep}(\text{Pt})$.

The temperature evolutions in all the configurations of the Pt/Co/Pt trilayers are reliably described by the 3TM with a consistent set of parameters. The largest sensitivity to $g_{ep}(\text{Pt})$ is obtained when the optically thick Pt layer is directly excited by a pump laser pulse and the Co thermometer is separated from the irradiated surface farther than approximately the optical absorption depth of Pt. Thus, to determine $g_{ep}(\text{Ru})$, we use a similar configuration, i.e., Pt/Co/Pt/Ru structure where the Ru layer is 50-nm-thick and directly excited by the pump pulse.

The temperature evolutions of the two transition metals, Pt and Ru, show distinct two-step temperature changes when the films are optically thick. We compare the results with the predicted temperatures in noble metals by taking Cu as an example when only purely diffusive transport is assumed. The properties of electrons and phonons as well as the aspects of thermalization processes in transition metals are discussed in comparison with noble metals.

## II. EXPERIMENTAL METHODS

We prepare samples by DC magnetron sputtering at base pressures below $5\times10^{-8}$ Torr. The sample stacks from top to bottom are Pt(2)/Co(0.8)/Pt(4), Pt(42)/Co(0.8)/Pt(4), Pt(16)/Co(0.8)/Pt(24), and Pt(2)/Co(0.8)/Pt(2)/Ru(50) on *c*-cut sapphire substrates, where the numbers in parentheses represent the layer thicknesses in nm. The thicknesses of the layers are determined by X-ray reflectivity measurements (X'pert, Philips). All the samples have perpendicular magnetic anisotropy as confirmed by a vibrating-sample magnetometer (MPMS, Quantum Design).

Pump-probe measurements are performed using a Ti:sapphire laser with wavelength centered at 785 nm. The train of pulses is generated at 80 MHz and split into two orthogonally polarized beams, i.e., pump and probe beams, by a polarizing beam splitter. The pump beam is modulated at 10.7 MHz by an electro-optic modulator. The probe beam is time-delayed relative to the pump and modulated by a chopper at 200 Hz. The $1/e^2$ beam radius ($w_0$) of the convolution of the pump and probe beams is 5.5 μm. The two-tint scheme [23] is employed to separate the wavelength spectrum of the pump and probe beams. For TDTR measurement, the intensity of the reflected probe is measured by a Si photodetector. For TR-MOKE measurements, the Kerr rotation of the reflected probe is measured by a combination of a half-wave-plate, a Wollaston prism, and a balanced photodetector. The voltage output of the Si detector or of the balanced detector is connected to an RF lock-in amplifier synchronized to the modulation frequency of the pump. To improve the signal-to-noise ratio, data are averaged over 10–15 repetitions when the pump and probe are incident on the opposite directions. The noise level is $0.2$ μrad$/\sqrt{N}$ with the time constant of 0.7 seconds where *N* is the number of repetitions. The TR-MOKE is taken as the difference of the MOKE signals of Co at remanence of the opposite polarities.

The pulse duration of the cross-correlation of the pump and probe beams is measured via two-photon absorption in a GaP photodetector. The zero-time delay when the pump and probe are incident on the opposite surfaces of a sample is determined by using the inverse Faraday effect of Pt, see Appendix B.

## III. RESULTS

Figure 1 shows TR-MOKE and TDTR measurement results for Pt(2)/Co(0.8)/Pt(4)/sapphire with the pump and probe incident on the Pt(2) surface. The absorbed pump fluence is 0.4 J m$^{-2}$ and the relative absorbance of each layer is calculated using a transfer matrix method, see Table S1. The temperatures of electrons, magnons, and phonons as a function of time and position are calculated using the 3TM in Eqs. (1-3) with $g_{ep}$(Pt) and thermalization time of electrons and magnons, $\tau_{em}$(Co)=$C_m$(Co)/$g_{em}$(Co) as two free parameters. See Table 1 for the materials parameters used in the model. The measured data are normalized to the calculated $\Delta T_m$(Co) at the delay time of 50 ps based on the absorption profile and the known thermal properties of substrate and the metal/substrate interface. The full-width-at-half-maximum (FWHM) of the cross-correlation of the pump and probe pulses is 1 ps and the zero-time-delay is set at the peak of the cross-correlation of pump and probe pulses.

We compare the TR-MOKE data with the magnon temperature of Co as the sample stays in the linear response regime and far below the Curie temperature. [21,22] The comparison between the TDTR data and the calculated electron and phonon temperatures is complicated at short delay times, < 5 ps, because the near-surface region of the sample is subject to temperature and strain gradients that contribute to the changes in optical reflectivity. A calculation of the reflectance change, $\Delta R$, requires knowledge of the derivatives of the complex optical index of refraction $\tilde{n}$ with respect to temperature $T$ and strain $\varepsilon$, i.e.,

$d\tilde{n}/dT = d\tilde{n}/dT_e + d\tilde{n}/dT_{ph} + d\tilde{n}/d\varepsilon$, which are difficult to obtain. In Fig. 1(a-b), we plot the calculated $\Delta R$ by considering $d\tilde{n}/dT$ [24] due to $T_{ph}(z, t)$ only. The simulated $\Delta R(T_{ph})$ captures the TDTR data only after electrons and phonons are equilibrated. Therefore, without specific information of $d\tilde{n}/dT_i$ ($i=e, ph$) and $d\tilde{n}/d\varepsilon$, accurate separation of electron and phonon temperatures from TDTR data remains challenging. Therefore, we use only the TR-MOKE data for quantitative analysis of non-equilibrium dynamics of our samples.

Upon laser excitation, the calculated electron temperature ($T_e$) rises sharply, followed by the rise of the magnon temperature ($T_m$). Electrons, magnons, and phonons are equilibrated at about 5 ps, as can be seen in Fig. 1(a-b). The temperature evolutions at these short delay times are dominated by energy exchange processes between the heat carriers. Heat transport is unimportant for this sample because the entire Pt/Co/Pt trilayer is directly excited by the laser pulse. The plateau of the temperatures until time delay of ≈50 ps indicates that heat is still confined in the metallic Pt/Co/Pt layers. After 50 ps, heat is transferred from the metallic layers into the dielectric substrate and the temperature evolution is primarily determined by phonon properties: phonon heat capacities ($C_{ph}$) of the metallic layers and the interface thermal conductance of phonons ($G_{ph}$) between Pt and sapphire substrate, $G_{ph} \approx 110$ MW m$^{-2}$ K$^{-1}$.

The sensitivity of the magnon temperature change ($\Delta T_m$) to a material parameter ($\alpha$) is defined as follows and shown in Fig. 1(c). $\Delta T_{m,max}$ is the maximum temperature change of magnons.

$$S(\alpha) = \frac{\partial(\Delta T_m)/\Delta T_{m,\max}}{d\alpha/\alpha} \tag{5}$$

The initial temperature rise at time delays ≤ 5 ps is most sensitive to $g_{ep}$(Pt), one of the free parameters in the model. For the electronic heat capacity coefficient of Pt, $\gamma_e$(Pt)=$C_e$(Pt)/$T$, we use the value from literatures, see Appendix C. The carrier coupling properties of Co, i.e., $\tau_{em}$(Co)=$C_m/g_{em}$, and $g_{ep}$(Co), affect the position of the temperature peak. We previously reported $\tau_{em}$(Co)≈0.2 ps and $g_{ep}$(Co)≈2×10$^{18}$ W m$^{-3}$ K$^{-1}$ by using time-resolved quadratic MOKE on 10-nm-thick Co. [22] We use the values for Co from Ref. [22] but include $\tau_{em}$(Co) of the 0.8-nm-thick Co layer as another free parameter. The sensitivity to *absorption depth* represents how sensitive the magnon temperature is to the optical absorption profile and is negligible in this optically thin sample. This absorption depth becomes important for the other optically thick samples.

To explore the effect of heat transport on temperature evolutions, we study samples with Pt layers thicker than the optical absorption depth of Pt, 11 nm. Figure 2(a) shows the results of the TR-MOKE and TDTR measurements of Pt(42)/Co(0.8)/Pt(4)/sapphire, when the pump and probe beams are incident through the transparent sapphire substrate and on Pt(4) surface with the absorbed fluence of 0.7 J m$^{-2}$. As the Co layer directly absorbs 3.5% of the total fluence (see Table S1), $T_e$ and $T_m$ of Co increase up to 30 K across the zero-time-delay and cool quickly and thermalize with phonons at ≈ 5 ps. The energy is then distributed within the Pt/Co/Pt layers.

Figure 2(b) shows the results for the same sample but with the pump beam incident on the top Pt(42) surface with the absorbed fluence of 1.2 J m$^{-2}$. The Co layer absorbs only 0.2% of the total fluence (see Table S1) and is mainly heated via heat transport into the Co layer from the adjacent region of the irradiated Pt layer. The initial temperature rises of electrons and magnons in Co are significantly reduced and followed by the second temperature rise at delay times > 5 ps. For the temperature evolutions at short delay times < 5 ps plotted on linear axes, see Fig. S1.

The temperature evolutions in Pt(42)/Co(0.8)/Pt(4) induced by front- or back-side heating of a pump pulse, calculated by the 3TM for the same absorbed fluence of 0.7 J m$^{-2}$, are shown in Fig. 2(c). The initial temperature evolutions are drastically different for the two geometries, but they eventually follow a common behavior at time delays > 50 ps. This is when the Pt/Co/Pt layers are thermalized and at time delays > 100 ps, heat transport into the sapphire substrate occurs. See Fig. S2 for the temperature evolutions as a function of depth at fixed delay times of 0 and 50 ps.

In the optically thick samples, heat transport across the Pt layer significantly affects the temperature evolutions. The sensitivities to the materials parameters are shown in Fig. 3. The temperature change of magnons is still most sensitive to $g_{ep}$(Pt) and has reduced sensitivities to the properties of Co as the Co layer is much thinner than the Pt layers. The sensitivity to electron thermal conductivity of Pt, $\Lambda_e$(Pt), is now larger and has opposite signs for the front- and back-side heating. Since our thermal model assumes purely diffusive transport, we expect a systematic deviation between the measured data and model if the heat carriers are not primarily diffusive and ballistic transport is significant. The electron thermal conductance at the Pt/Co interface, $G_{ee}$(Pt/Co), also affects heat transport, but the value is large, i.e., $\geq 8$ GW m$^{-2}$ K$^{-1}$, and has a negligible effect on $\Delta T_m$(Co). See Fig. S3 for determination of $G_{ee}$(Pt/Co) and Fig. S4 for parameters that have large sensitivities but are known or determined by other methods.

Another important parameter in the optically thick samples is the *absorption depth*, as shown in Fig. 3. Note that "*absorption depth*" is an effective parameter that we use to approximate the optical absorption profile across the metallic layers as a single exponential curve and only used for calculating the sensitivity. The large sensitivity to the *absorption depth* at time delays < 2 ps implies that direct optical absorption significantly contributes to the initial temperature evolutions, even when the amount of absorbed energy is not large, e.g., 0.2% for the Co layer below the 42-nm-thick Pt layer. See Appendix D for further discussion of relative contributions of direct optical absorption and electron heat transport to initial temperature evolutions in Co.

TR-MOKE data of all the samples are shown in Fig. 4 for delay times < 5 ps. In addition to the Pt(42)/Co(0.8)/Pt(4) sample, we measure Pt(16)/Co(0.8)/Pt(24)/sapphire, where the Co layer is located approximately in the middle of the optically thick Pt layer and heated from either front or back surface. (The measurement results of Pt(16)/Co(0.8)/Pt(24) are shown in Fig. S5.) The Co layers in the Pt samples of the five different configurations demagnetize almost at the same rate and peak at (0.6±0.1) ps. The peak positions do not shift significantly as the Co layer is located farther from the irradiated Pt surface. This observation is in contrast to Ref. [25], in which the demagnetization peak of a [Co/Pt] multilayer is delayed as the Cu layer subject to laser irradiation gets thicker. We discuss the different mechanisms for initial non-equilibrium dynamics in Pt and Cu in Sec. IV.

The uncertainties in the two free parameters, $g_{ep}$(Pt) and $\tau_{em}$(Co), are evaluated with the criterion of $\sigma=2\sigma_{min}$, where $\sigma$ is the sum of square of the residuals between the 3TM calculation and TR-MOKE data. The contours for $\sigma=2\sigma_{min}$ for all the samples are shown in Fig. 5. We obtain $g_{ep}$(Pt)=(6±1)×10$^{17}$ W m$^{-3}$ K$^{-1}$ and $\tau_{em}$(Co)=(0.23±0.05) ps that are consistent in all the five configurations of the Pt/Co/Pt trilayers. We do not observe any systematic deviations between the model and data. If ballistic transport of hot electrons is dominant over diffusive transport, the electron temperature in the Co layer would be higher than the 3TM prediction when the Co layer is farther from the irradiated Pt layer, which would result in a lower value of apparent $g_{ep}$(Pt). However, we do not see such a deviation. This can be attributed to the fact that the bulk mean-free-paths of electrons near the Fermi energy in Pt are about 7-10 nm at room temperature, [20] and shorter than or comparable to the length scale of temperature gradient, i.e., optical absorption depth.

Prior reports for $g_{ep}$(Pt) are relatively few compared with those of noble metals: The authors of Ref. [2,26] used TDTR of Pt thin films and determined $g_{ep}$(Pt) as 2.5 and $10.9 \times 10^{17}$ W m$^{-3}$ K$^{-1}$, respectively, by assuming $\Delta R \propto \Delta T_e$; The authors of Ref. [27,28] used TDTR signals from the Cu side of Pt(20)/FM(3)/Cu(100) (FM=ferromagnetic multilayer) samples and derived $g_{ep}$ =4.2 and $2.9 \times 10^{17}$ W m$^{-3}$ K$^{-1}$, respectively, by assuming $\Delta R \propto \Delta T_{ph}$; Ref. [29] used time-resolved photoemission spectroscopy of Pt and derived $g_{ep}$(Pt)=$6.8 \times 10^{17}$ W m$^{-3}$ K$^{-1}$ at 77 K. All of the prior work uses an electronic heat capacity derived from the linear temperature coefficient determined from an extrapolation of the low-temperature measurements of bulk Pt to room temperature, $\gamma_e$(Pt) $\approx$ 720 J m$^{-3}$ K$^{-2}$ [30]. We use $\gamma_e$(Pt) = 400 J m$^{-3}$ K$^{-2}$ from the first-principles calculations and the temperature-dependence of experimental heat capacity of Pt, see Appendix C.

We apply the Pt/Co/Pt trilayer as a thermometer to determine the carrier coupling parameter in Ru using the structure Pt(2)/Co(0.8)/Pt(2)/Ru(50)/sapphire. The highest sensitivity to $g_{ep}$ of the metal layer is obtained when the irradiated metal layer is optically thick and the Co layer is separated from the irradiated surface but close enough to exhibit the initial temperature rise. This is the geometry where the pump is incident on the optically thick Ru surface and the probe is on the other surface. (See Fig. S6) The overall temperature evolutions resemble those in Pt(42)/Co/Pt(4) with front-side heating in Fig. 2(b). As a result, we obtain $g_{ep}$(Ru)=$(9\pm2) \times 10^{17}$ W m$^{-3}$ K$^{-1}$. For comparison, Ref. [31] reported $g_{ep}$(Ru)=$18.5 \times 10^{17}$ W m$^{-3}$ K$^{-1}$ by measuring thermoreflectance using the pump-pump-probe technique.

## IV. DISCUSSION

An important metric that describes non-equilibrium between electrons and phonons is the length scale, $l_{ep}=(\Lambda_e/g_{ep})^{1/2}$, which represents the characteristic distance over which electrons and phonons stay out-of-equilibrium. Based on the results of this work, this length-scale is ≈9 nm for Pt and ≈7nm for Ru. These values are even smaller than the optical absorption depths, 11 nm and 13 nm for Pt and Ru, respectively. On the other hand, noble metals, e.g., Au, Ag, and Cu, have larger $\Lambda_e$ and smaller $g_{ep}$ than transition metals, and therefore have much longer $l_{ep}$. For example, Cu has $l_{ep}$ ≈63 nm, much longer than its absorption depth, 8 nm. (See Table 1) This indicates that electrons in Cu can maintain a higher temperature than phonons over the long length scale of $l_{ep}$.

To illustrate the effect of $l_{ep}$ on temperature evolutions, we compare the simulated temperatures of Cu and Pt by using the 2TM considering only diffusive transport, see Fig. 6. We use the parameters for Cu from Ref. [25] and for Pt from this work, as shown in Table 1. Figure 6(a, c) shows the electron temperature at the bottom of the metal single layer on sapphire substrates when the top surface is irradiated with a 50-fs-long laser pulse. The peak appears almost identical for the Pt layer regardless of its thickness (Fig. 6(a)), similar to what we observe for the Pt/Co/Pt layers shown in Fig. 4. On the other hand, the initial temperature peak is delayed as the Cu layer gets thicker (Fig. 6(c)). The faster temperature rise in Pt than in Cu seems counter-intuitive given the higher electron diffusivity of Cu than Pt, i.e., $D_e=\Lambda_e/C_e$≈$10^{-2}$ m² s$^{-1}$ and $4\times10^{-4}$ m² s$^{-1}$, respectively. In fact, these different behaviors are because heat transport occurs differently in Pt and Cu.

Figure 6(b, d) shows electron and phonon temperatures at the bottom surface of 100-nm-thick of Pt and Cu single layers on sapphire. In Fig. 6(b), Pt shows a small temperature rise of electrons across the zero-time-delay, followed by a larger increase in temperature at time delays

> 1 ps when electrons and phonons are reaching thermal equilibrium. This type of temperature evolution is observed in our experiments of the optically thick layers of Pt in Fig. 2(b) and Ru in Fig. S6. The initial temperature rise is caused by both direct optical absorption and electron heat transport from Pt, as we discuss in Appendix D. We expect that electrons only within the length scale of $l_{ep}$ from the bottom surface contribute to the temperature change at the bottom surface, while electrons from the irradiated top surface would lose energy to phonons before reaching the bottom surface. This explains why the position of the peak temperature is independent of the thickness of Pt in Fig. 6(a). In Pt, electrons carry a limited amount of energy to the bottom surface before equilibrating with phonons, and most of the laser energy is distributed across the Pt layer by electrons and phonons in near thermal equilibrium.

On the other hand, electrons in Cu transfer most of the laser energy to the bottom surface within a few ps following laser excitation, while phonons remain cold. Then the electron temperature decreases due to energy transfer from electrons to phonons, followed by a plateau implying the heat carriers in a Cu layer are thermalized, as shown in Fig. 6(d). This is in stark contrast to the dynamics in Pt. Moreover, the temperature evolution in Cu shown in Fig. 6(b) shows that the travel time increases approximately linearly with the thickness of Cu, which has been considered in the past as evidence for ballistic transport. (See Appendix F for further discussion of the apparent linear relationship.) Therefore, it is difficult to assign a transport mechanism based on the functional form of the thickness dependence. The apparent dynamics are complicated due to the presence of various heat transfer processes, e.g., direct optical excitation, carrier interactions, and ballistic and diffusive transport.

Transport of hot electrons in non-magnetic metals is an important driver for ultrafast demagnetization of an adjacent ferromagnetic material. Although prior work emphasizes the role of ballistic transport of hot electrons in the non-magnetic metallic layer for inducing indirect excitation of the adjacent ferromagnetic layer, [25,32–34] we argue that based on the results of this work, diffusive transport of electrons can be as effective as ballistic transport in terms of heat flux when $l_{ep}$ is substantially longer than the optical absorption depth. Our work also gives a different interpretation of the results of Ferté et al. [33]: the authors showed the demagnetization of CoTb occurs more slowly when CoTb is heated by a Cu/Pt/Cu trilayer than by a single Cu layer of similar thicknesses. The authors attributed the difference in the demagnetization rate to the different velocities of diffusive and ballistic transport of electrons. However, the slow demagnetization rate may arise from the electron diffusivity in Pt by a factor of 25 smaller than the electron diffusivity of Cu, without requiring the concept of ballistic transport.

## V. SUMMARY

An ultrathin Co layer sandwiched by Pt layers exhibits perpendicular magnetic anisotropy and can serve as an effective thermometer for detecting non-equilibrium dynamics in adjacent metals via TR-MOKE. We demonstrate that the 3TM can adequately describe the temperature evolutions in the five different configurations of the Pt/Co/Pt trilayers with a consistent set of materials parameters, including $g_{ep}(\text{Pt})=(6\pm1)\times10^{17}$ W m$^{-3}$ K$^{-1}$ and $\tau_{em}(\text{Co})=C_m/g_{em}=(0.23\pm0.05)$ ps. The maximum sensitivity to $g_{ep}$ (Pt) is obtained when the optically thick Pt layer is irradiated with a laser pulse and the Co thermometer is located beyond the optical absorption depth of the Pt layer. Using the optimum configuration, we determine $g_{ep}(\text{Ru})=(9\pm2)\times10^{17}$ W m$^{-3}$ K$^{-1}$ using the Pt/Co/Pt/Ru structure where the Ru layer is 50-nm-thick. The initial temperature dynamics in Pt and Ru are governed by direct optical excitation and electron heat transport within the non-equilibrium length scale ($l_{ep}$) and most of the heat transport of the laser energy occurs while electrons and phonons are near thermal equilibrium.

This is in contrast to the Cu, where electron heat transport is distinct while the lattice stays cold.

Figure 1. Temperature evolutions in Pt(2)/Co(0.8)/Pt(4)/sapphire when pump is incident on the surface of the 2-nm-thick Pt film, i.e., Pt(2), with absorbed pump fluence of 0.4 J m$^{-2}$. Open symbols are time-resolved magneto-optic Kerr effect (TR-MOKE) (red) and time-domain thermoreflectance (TDTR) (black) measured with probe incident on the Pt(2) surface on (a) log-log axes and (b) linear axes at short delay times. Solid lines are the best-fit of electron (blue), magnon (red), and phonon (black) temperatures of Co calculated by a three-temperature model (3TM). Also shown is reflectance change ($\Delta R$) calculated using a spatial profile of phonon temperature (green). (c) Sensitivities of magnon temperature of Co to materials parameters. $g_{ep}$, $\tau_{em}$, and $\gamma_e$ are electron-phonon coupling parameter, electron-magnon thermalization time, and electron heat capacity coefficient, respectively.

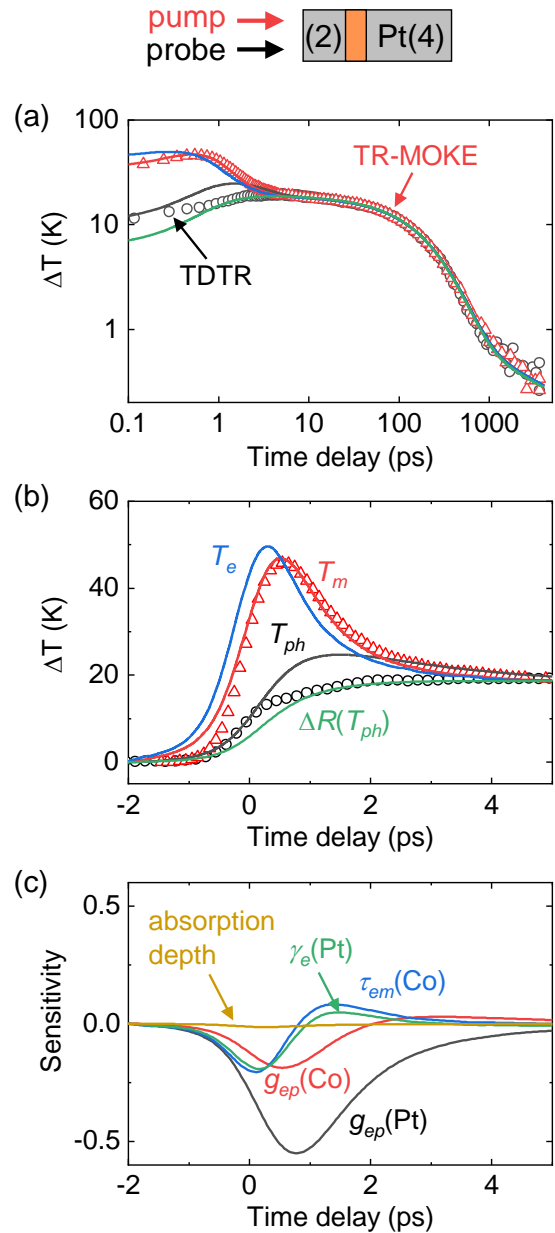

Figure 2. Temperature evolutions in Pt(42)/Co(0.8)/Pt(4)/sapphire when pump is incident on (a) Pt(4) surface and (b) Pt(42) surface. Open symbols are TR-MOKE (red) and TDTR (black) measured with probe incident on Pt(4) surface. Solid lines are the best fit of electron (blue), magnon (red), and phonon (black) temperatures of Co calculated by a 3TM. Also shown is reflectance change calculated using a spatial profile of phonon temperature (green). (c) Calculated temperatures for the sample configurations in (a) (solid lines, "back side") and (b) (dotted lines, "front side") with the same absorbed fluence of 0.7 J m$^{-2}$.

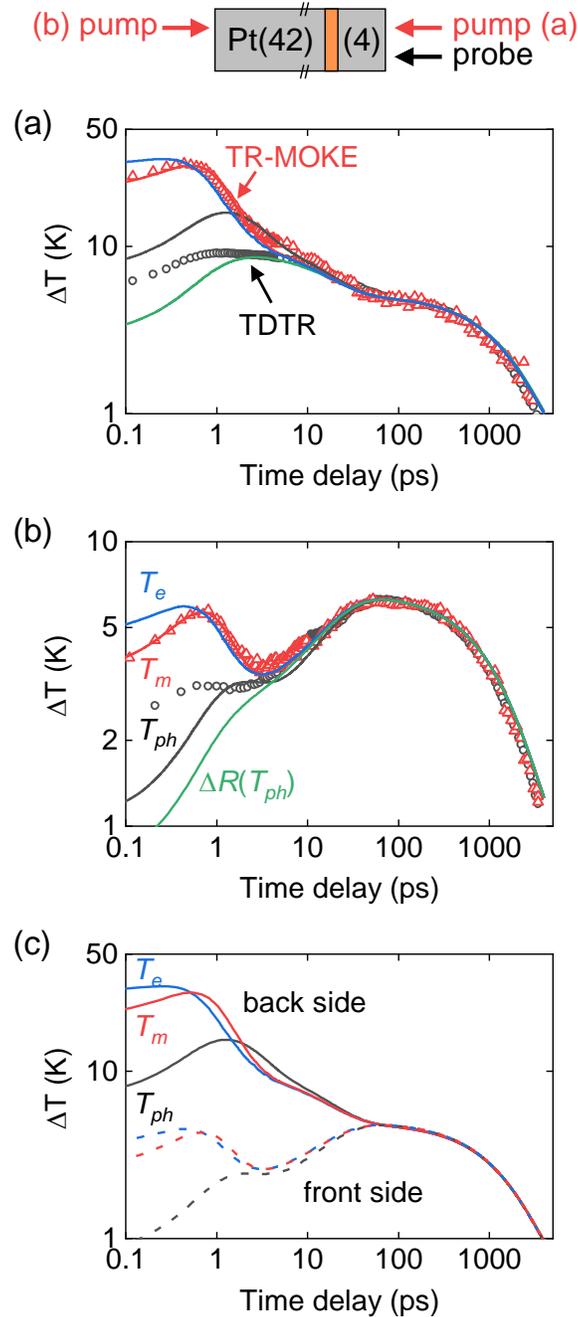

Figure 3. Sensitivities of magnon temperatures to materials parameters in Pt(42)/Co(0.8)/Pt(4)/sapphire when pump is incident on (a) Pt(4) surface and (b) Pt(42) surface. $\Lambda_e$ is electron thermal conductivity.

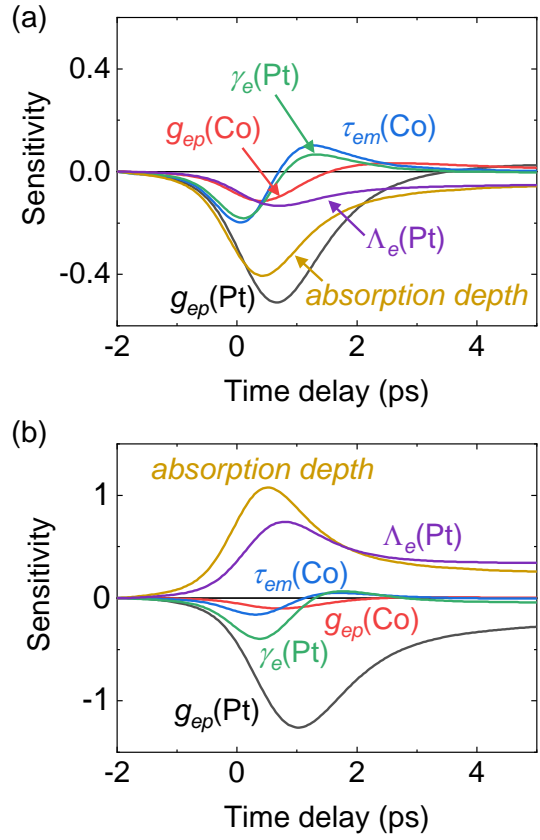

Figure 4. TR-MOKE at short delay times for five configurations of Pt/Co/Pt trilayers on sapphire substrates. "Front" or "back" indicates the surface on which the pump pulse is incident. The *y*-axis is the Kerr rotation normalized to its maximum change that occurs at time delays < 2 ps.

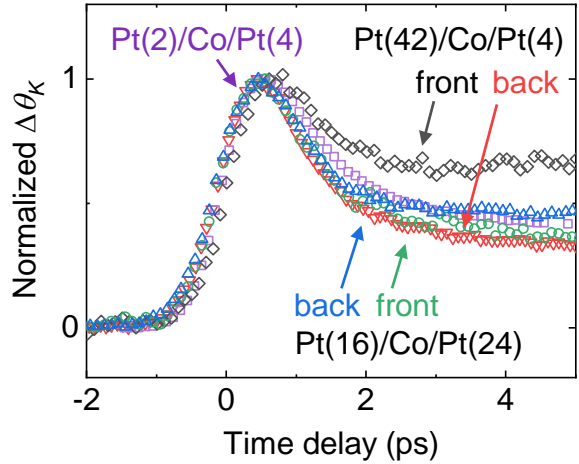

Figure 5. Best fit of electron-phonon coupling constant of Pt, $g_{ep}$(Pt), and electron-magnon thermalization time of Co, $\tau_{em}$(Co), for five configurations of Pt/Co/Pt trilayers on sapphire substrates. "Front" or "back" is the surface on which the pump pulse is incident.

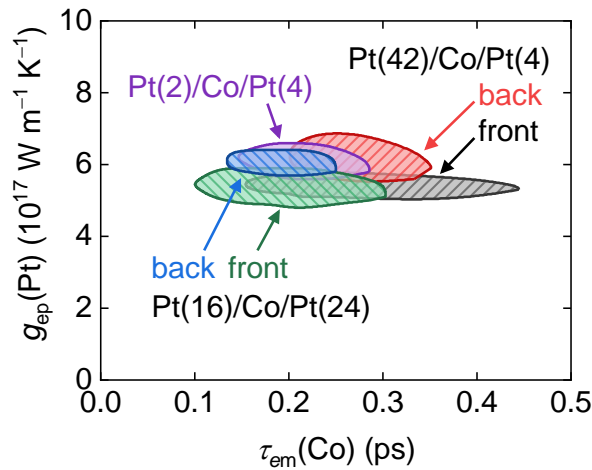

Figure 6. Temperature evolutions at the bottom surface of (a-b) Pt and (c-d) Cu single layers on sapphire substrates calculated by a 3TM when the top surface is irradiated with a laser pulse of 50 fs pulse duration. (a, c) Electron temperature changes are normalized to the maximum changes at time delays < 2 ps. The numbers represent the thickness of (a) Pt and (c) Cu layers. (b, d) Electron (black) and phonon (red) temperatures at the bottom of the (b) Pt and (d) Cu layers of 100 nm in thickness for the same absorbed fluence of 0.35 J m$^{-2}$.

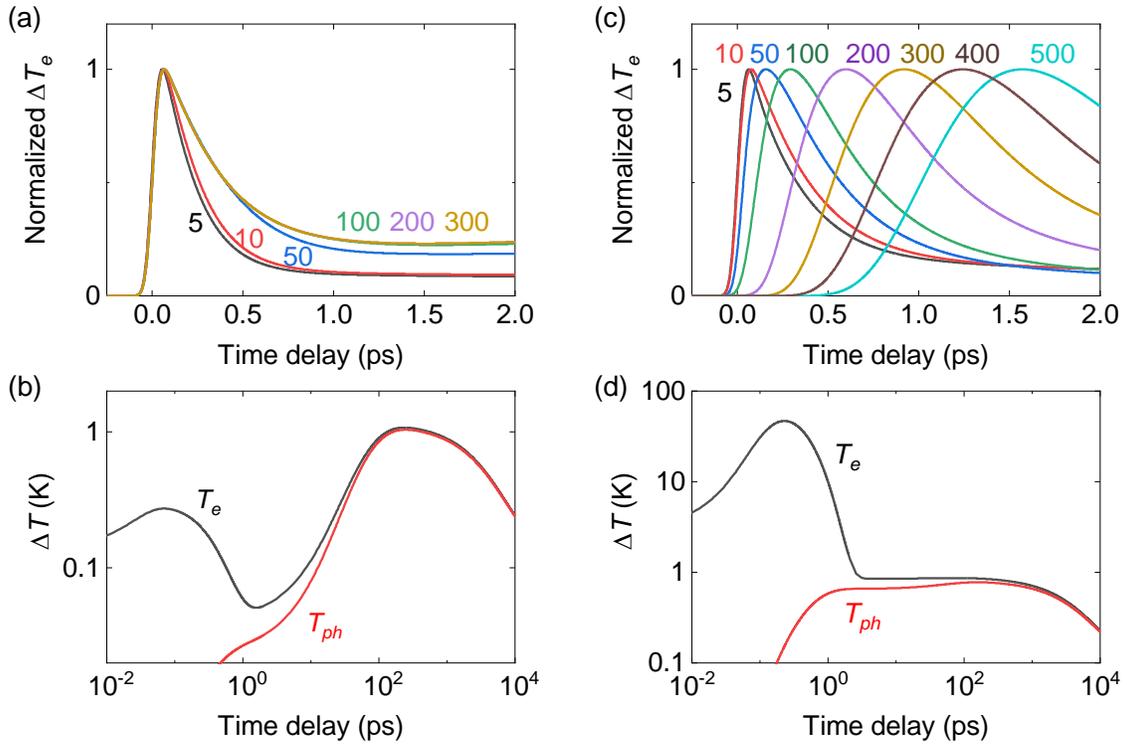

Figure 7. Determination of zero-time-delay via inverse Faraday effect. Open symbols are TR-MOKE data of Pt(16)/Co(0.8)/Pt(24)/sapphire when a circularly polarized pump beam is incident on Pt(24) surface, and a linearly polarized probe beam is incident on Pt(16) surface. The sum of TR-MOKE data for right- and left-circularly polarized pump beams ("RCP+LCP", blue symbol) represents magnetization dynamics in Co, and the difference ("RCP−LCP", black symbol) represents the inverse Faraday effect. Red solid line is the scaled intensity of a correlated pump-probe pulse measured via two-photon absorption.

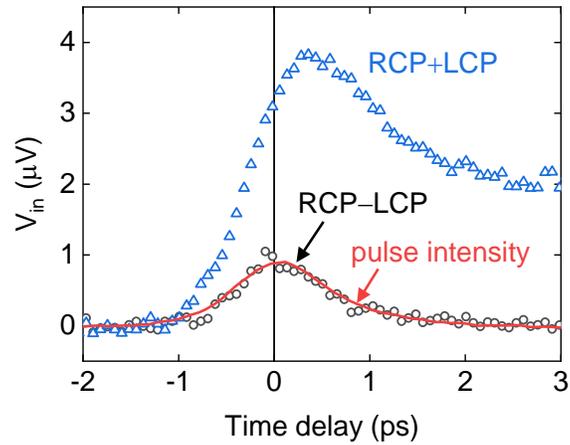

Figure 8. Onset-time ($t_{0.1}$) of electron temperature at the bottom surface of a Cu layer versus (a) thickness and (b) thickness-squared of the Cu layer. Black lines are the onset-times extracted from Fig. 6(a) and red lines are linear fits.

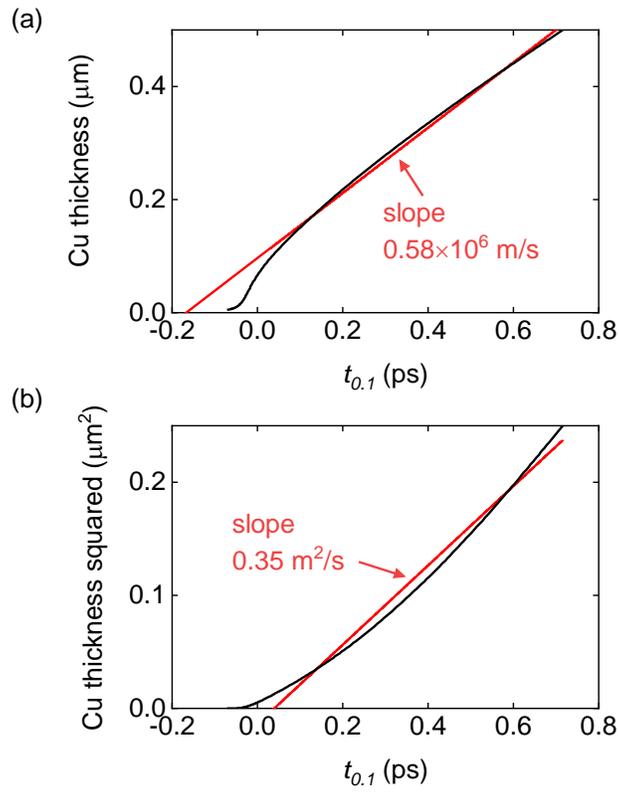

**TABLE 1.** Materials parameters that are used for calculating the three-temperature model.

| | Pt | Ru | Cu | Co | $Al_2O_3$ |
|---|---|---|---|---|---|
| $C_{total}$ ($10^6$ J m$^{-3}$ K$^{-1}$) | 2.82 [a] | 2.97 | 3.45 | 3.75 | 3.08 |
| $\gamma_e$ (J m$^{-3}$ K$^{-2}$) | 400 [b] | 371 [c] | 100 | 680 [c] | – |
| $C_m$ ($10^6$ J m$^{-3}$ K$^{-1}$) | – | – | – | 0.02 (bulk) [d] | – |
| $\Lambda_{ph}$ (W m$^{-1}$ K$^{-1}$) | 7 [e] | 7 [e] | 10 | 14 [f] | 33 |
| $\Lambda_e$ (W m$^{-1}$ K$^{-1}$) | 50 [g] | 50 [g] | 300 | 20 [g] | – |
| $g_{ep}$ (W m$^{-3}$ K$^{-1}$) | $(6\pm1)\times10^{17}$ | $(9\pm2)\times10^{17}$ | $7.5\times10^{16}$ | $(2.0\pm0.2)\times10^{18}$ | – |
| $l_{ep}=(\Lambda_e/g_{ep})^{1/2}$ (nm) | 9 | 7 | 63 | 3 | – |
| $\tau_{em}=C_m/g_{em}$ (ps) | – | – | – | $0.23\pm0.05$ | – |
| $\tilde{n}$ | 2.7+i5.9 [h] | 5.17 + i4.91 | 0.25+i5.03 | 2.5+i4.8 [i] | 1.76 |
| $d\tilde{n}/dT$ | $2.6\times10^{-4}+i(-3\times10^{-4})$ [j] | – | – | $2.6\times10^{-4}+i(-3\times10^{-4})$ [j] | – |

a. Ref. [30]
b. Ref. [35]
c. Ref. [36]
d. Ref. [37]
e. Ref. [38]; $\Lambda_{ph}$ of Ru is assumed to be the same as $\Lambda_{ph}$ of Pt.
f. $\Lambda_{ph}$ of Co is assumed to be the same as $\Lambda_{ph}$ of Ni in Ref. [39].
g. Electrical conductivities of Pt, Ru, and Co are measured by using a four-probe method on Pt and Ru single layers of 50 nm in thickness and a 10-nm-thick Co layer capped with Pt 2 nm on sapphire substrates. The electronic thermal conductivities are derived via the Wiedemann-Franz law.
h. Ref. [40]
i. Ref. [41]
j. Ref. [24]. The $d\tilde{n}/dT$ of Co is not available and assumed to be the same as that of Pt as the thermoreflectances, $dR/dT$, of Co and Pt are the same within the experimental uncertainty for 785 nm wavelength. [24]

**TABLE 2.** Calculated energy per area that is stored in the Co layer in Pt($h$)/Co(0.8)/Pt(5)/sapphire from electron heat transport from Pt layers and direct optical absorption in the Co layer. Heat flux from top/bottom Pt layers into the Co layer is estimated as integration of $G_{ee}\Delta T_e$ over delay time up to 1 ps. Total energy into Co is the sum of the energy per area of the heat flux and optical absorption. We assume a constant total absorbed fluence of 0.5 J m$^{-2}$ and a pulse duration of 1 ps FWHM. (See Appendix D)

| $h$ (nm) | Heat flux from top Pt (%) | Heat flux from bottom Pt (%) | Absorption in Co (%) | Total energy into Co (mJ m$^{-2}$) | Max $\Delta T_m$(Co) (K) |
|---|---|---|---|---|---|
| 10 | 66  | 4   | 31 | 56 | 30.3 |
| 20 | 96  | −22 | 26 | 29 | 15.6 |
| 30 | 118 | −40 | 22 | 14 | 7.7  |
| 40 | 132 | −52 | 19 | 6  | 3.5  |

## APPENDIX

### A. Lifetimes of electronic excitations in Pt

Fermi liquid theory provides an estimate for the lifetimes of hot electrons, $\tau_{ee}(E)$, within the random phase approximation [9]

$$\frac{1}{\tau_{ee}(E)} = \frac{1}{\tau_0} \frac{(E-E_F)^2}{E_F^2} \quad (A1)$$

where $\tau_0 = \frac{128}{\pi^2 \sqrt{3}} \frac{1}{\omega_p}$ and $\omega_p$ is plasma frequency. Eq. (1) is based on the free-electron gas model and is a good approximation for intra-band transitions in simple metals. If the relaxation of hot electrons follows Fermi-liquid theory, $\tau_{ee}(E)[E-E_F]^2$ is constant and equal to $\tau_0 E_F^2$.

Time-resolved photoemission spectroscopy [9] provides measurements of $\tau_{ee}(E)$. Noble metals show qualitative agreement with Fermi-liquid theory. While Eq. (1) predicts $\tau_0 E_F^2 \approx 30$ and 17 fs eV$^2$ for Cu and Au, respectively, photoemission spectroscopy measurements give $\tau_{ee}(E)[E-E_F]^2 \approx 45$ and 75 fs eV$^2$ for Cu and Au, respectively, in the energy range of 0.5 eV $\leq$ $E$-$E_F \leq$ 2 eV. The disagreement between $\tau_{ee}(E)[E-E_F]^2$ predicted by Eq. (A1) and the photoemission spectroscopy data for Cu and Au can be attributed to the screening effect of completely filled $d$ bands lying >2 eV below the Fermi level, which is more effective in Au than Cu. [9]

For Pt, photoemission spectroscopy data is unavailable to the best of our knowledge, but first-principles calculations [42] incorporating the full band structure suggest that $\tau_{ee}(E)[E-E_F]^2$ deviates from Fermi-liquid theory and strongly depends on excitation energy, i.e., sharply increases with increasing excitation energy up to 3 eV with $\tau_{ee} \approx 5$ fs at $(E–E_F) = 1$ eV. Insight about the behavior of Pt can be drawn from experimental data for Pd as both Pd and Pt have similar valence electronic structures. Photoemission data for $\tau_{ee}(E)$ of Pd [9] shows similar energy-dependence and magnitude to those of the theoretical prediction of $\tau_{ee}(E)$ for Pt. [42]

Additional experimental insights on $\tau_{ee}(E)$ can be drawn from the electrical resistivity measurements at $T < 20$ K [43], where electron-electron scattering dominates over electron-phonon scattering. The electrical resistivity due to electron-electron scattering ($\rho_{ee}$) exhibits a $T^2$ dependence; extrapolation to 300 K gives $\rho_{ee} \approx 1$ μΩ cm, ≈10% of the total resistivity. According to Ref. [44], the product of $\rho\tau$ is determined by the shape of the Fermi surface and gives $\tau_{ee}$ of Pt at 300 K ≈ 100 fs. If we approximate the excitation energy of near-equilibrium conduction electrons at room temperature as $2k_BT = 0.05$ eV, $\tau_{ee}(E)[E - E_F]^2$ is ≈ 0.25 fs eV$^2$ and $\tau_{ee}(E) \approx 100$ fs. Thus, we estimate that the electron-electron thermalization timescale in Pt is more than an order of magnitude shorter than in Cu or Au.

## B. Pulse duration and zero-time delay

The pulse duration of the cross correlation of pump and probe pulses is measured via two-photon absorption using a GaP detector. The measured intensity profile approximately follows a Gaussian shape with the full-width-at-half-maximum (FWHM) of 1 ps. This limits the time resolution of our measurements to 1 ps. The laser pulses are strongly stretched in time at the sample position compared to the output of the laser due to significant dispersions in the electro-optic modulator and sharp-edge optical filters. In our experiments, the response of the system is linear to a good approximation, therefore, only the convolution of the pump and probe pulses matters and is used as an input for $P(t)$ in Eq. (4).

The zero-time-delay when the pump and probe are incident on the same surface of the samples can be determined using the GaP detector. The zero-time-delay when the pump and probe are incident on the opposite sides of the samples is determined via the inverse Faraday effect. [45] The pump beam is circularly-polarized by a quarter-wave-plate and creates circular birefringence upon incidence on the sample surface; the lifetime of this birefringence is set by the momentum scattering time of the excited electrons. If the Pt layer that the pump pulse is incident on is sufficiently optically transparent, the transient birefringence can be detected by the probe beam. The TR-MOKE signal then displays a narrow peak across the zero-time-delay, as shown in Fig. 7. For thicker Pt layers, e.g., Pt 42 nm, the nonlinear response of the detector generated by the leaked pump in the absence of the optical filter is used to determine the zero-time-delay. We correct the zero-time-delay if needed to take into account insertion or removal of various optical elements from the pump or probe beam paths. We estimate that the uncertainty in the position of zero-time-delay is approximately 0.1 ps.

## C. Electronic heat capacity coefficient of Pt, $\gamma_e$(Pt)

The electronic heat capacity coefficient, $\gamma_e$(Pt)=$C_e$(Pt)/$T$, is an important parameter for calculating the initial temperature rise of Pt electrons. Lin *et al.* [35] point out that $\gamma_e$ of metals is not a constant and depends on temperature, especially when the electronic density of states is not a smooth function of energy. First-principles calculation of the electron density of states of Pt [35] show that $\gamma_e$(Pt) decreases as temperature increases, and is ≈400 J m$^{-3}$ K$^{-2}$ at 300 K. [35]

The experimental heat capacity of Pt, $C_P(T)$, also suggests that $\gamma_e$(Pt) varies as a function of temperature in the range of 300 K ≤$T$≤ 2000 K. [30] As the Debye temperature of Pt is 236 K, the temperature-dependence of $C_P(T)$ in this temperature range is predominantly determined by the electronic heat capacity and the difference of $C_P$ and $C_V$, i.e., $C_P-C_V = \alpha^2 V_m T/\kappa_T$, where $\alpha$, $V_m$, and $\kappa_T$ are the thermal expansion coefficient, molar volume, and isothermal compressibility, respectively. After subtracting the thermodynamic term from $C_P/T$ at 300 K using the experimental values of $\alpha=2.75\times10^{-5}$ K$^{-1}$ [46], $V_m=9.09$ cm$^3$ mol$^{-1}$ [46], and $\kappa_T=3.26\times10^{-3}$ GPa$^{-1}$ [47], we obtain $\gamma_e \approx 410$ J m$^{-3}$ K$^{-2}$, similar to the result of the first-principles calculation [35]. Therefore, we fix $\gamma_e$(Pt)=400 J m$^{-3}$ K$^{-2}$ in our work. This gives $C_e$(Pt)=$\gamma_e T$=0.12×10$^6$ J m$^{-3}$ K$^{-1}$ out of the total heat capacity of Pt, 2.82 ×10$^6$ J m$^{-3}$ K$^{-1}$ at 300 K.

### D. Relative energy per area into Co layer in Pt/Co/Pt

The initial temperature change at the bottom of the optically thick Pt layer has two sources: direct optical excitation and electron heat transport. To evaluate the relative contributions of the two sources, we compare direct optical absorption in the Co layer and heat flux from the Pt layers into the Co layer in terms of energy per unit area for Pt($h$)/Co(0.8)/Pt(5)/sapphire structures where $h$=10–40 nm using the 3TM, see Table 2. We assume the constant total absorbed fluence is 0.5 J m$^{-2}$ and the FWHM of the optical pulse is 1 ps. The relative absorption in the Co layer is estimated from a transfer matrix method. The heat flux across the Pt/Co interface can be calculated as $J_q=G_{ee}\Delta T_e$, where $G_{ee}$ is the thermal conductance of electrons across the Pt/Co interface and $\Delta T_e$ is the difference of electron temperature at the interface. We estimate $G_{ee} \geq 8$ GW m$^{-2}$ K$^{-1}$ from TDTR measurement, as shown in Fig. S3. To obtain the energy that enters the Co layer from top and bottom Pt layers, we integrate the heat flux until the time delay of 1 ps.

In Table 2, the total energy into the Co layer represents the sum of the energy from heat flux and direct optical absorption in the unit of J m$^{-2}$. The heat flux and optical absorption are presented as relative fractions. The ratio of the maximum change of magnon temperature in Co to the total energy stored in Co appears consistent at ≈0.545 K m$^3$ J$^{-1}$. We note that heat flux can be underestimated as we use the minimum value of $G_{ee}$. These results show that both electron heat transport and direct optical absorption significantly contribute to the initial dynamics in the Co in Pt/Co/Pt.

**E. Apparent linear relationship between layer thickness and travel time**

The hallmark of ballistic transport has been considered as the travel time being linear with distance. Prior work supported the presence of ballistic transport by showing the onset-time of temperature rise is linear with the thickness of Au [13] and Cu [25] layers. However, we point out here that an apparent linear relationship alone does not conclusively establish the importance of ballistic transport. Figure 8 shows plots of the thickness and thickness-squared of Cu versus the calculated onset-time ($t_{0.1}$) extracted from Fig. 6(a). Only diffusive transport is considered in these calculations. The onset-time is defined as the time-delay where the temperature rise is 10% of the maximum temperature at the bottom surface of the Cu layer according to Ref. [25]. In Fig. 8(a), the thickness is linear with the onset-time and the apparent slope, 0.58×10$^6$ m s$^{-1}$, is close to the velocity of hot electrons reported in Ref. [25], 0.68×10$^6$ m s$^{-1}$. On the other hand, the slope in Fig. 8(b), 0.35 m$^2$ s$^{-1}$ is more than an order of magnitude greater than the electron diffusivity of Cu, $D_e$=10$^{-2}$ m$^2$ s$^{-1}$. Therefore, the fact that a model based on purely diffusive transport can produce a linear relationship between the thickness and travel time implies that one cannot conclusively determine the transport mechanism based the functional form alone.

**Acknowledgements**

The authors thank Dr. Zhu Diao for helpful discussions. Pump-probe measurements and sample analysis were carried in the Frederick Seitz Materials Research Laboratory Central Research Facilities, University of Illinois, and were supported by MURI W911NF-14-1-0016.